# Electrically switchable Casimir forces using transparent conductive oxides

Tao Gong[1,2], Benjamin Spreng[1], Miguel Camacho[3], Inigo Liberal[4], Nader Engheta[5] and Jeremy N. Munday[1]

[1]*Department of Electrical and Computer Engineering, University of California, Davis, USA*
[2]*Department of Materials Science and Engineering, University of California, Davis, USA*
[3]*Department of Electronics and Electromagnetism, University of Seville, Spain*
[4]*Institute of Smart Cities, Public University of Navarre, Spain*
[5]*Department of Electrical and Systems Engineering, University of Pennsylvania, USA*

**Abstract**
Casimir forces between charge-neutral bodies originate from quantum vacuum fluctuations of electromagnetic fields, which exhibit a critical dependence on material's electromagnetic properties. Over the years, *in-situ* modulation of material's optical properties has been enabled through various means and has been widely exploited in a plethora of applications such as electro-optical modulation, transient color generation, bio- or chemical sensing, etc. Yet Casimir force modulation has been hindered by difficulty in achieving high modulation signals due to the broadband nature of the Casimir interaction. Here we propose and investigate two configurations that allow for *in-situ* modulation of Casimir forces through electrical gating of a metal-insulator-semiconductor (MIS) junction comprised of transparent conductive oxide (TCO) materials. By switching the gate voltage on and off, a force modulation of > 400 pN is predicted due to substantive charge carrier accumulation in the TCO layer, which can be easily measured using state-of-the-art force measurement techniques in an atomic force microscope (AFM). We further examine the influence of the oxide layer thickness on the force modulation, suggesting the importance of the fine control of the oxide layer deposition. Our work provides a promising pathway for modulating the Casimir effect *in-situ* with experimentally measurable force contrast.

**Introduction**
Quantum vacuum fluctuations of electromagnetic fields are a fascinating quantum-mechanical effect, manifested by a multitude of celebrated physical phenomena such as Lamb shift, spontaneous emission, the anomalous magnetic moment of electron and surface wetting [1]. Amongst them is the Casimir effect, named after H. B. G. Casimir who, in 1948, predicted an attractive force between two perfectly conducting parallel plates that scales with the plate-plate separation $d$ as $\propto d^{-4}$ [2]. This force acting between two charge-neutral bodies results from the perfectly reflective boundary condition imposed on the fluctuating vacuum modes. The Casimir force is of fundamental research interest and has also brought about significant implications in nanotechnology [3-6].

If the perfectly reflective boundary condition is lifted by taking into account the optical properties of the materials comprising the interacting bodies (and the intervening medium), the magnitude and/or the sign of the Casimir force can be dramatically altered [7-16]. For instance, it has been demonstrated that the Casimir force between an Au-coated sphere and a transparent conductive oxide (TCO) film is nearly half of the value found between two noble metal films [9,17]. On the other hand, the force between an Au sphere and a silica film immersed in certain liquid solutions (e.g., Bromobenzene) were measured to be repulsive [11].

The material dependence of the Casimir force has also provoked the pursuit of direct force modulation by modifying the optical properties of the materials. Appropriate doping in semiconductors can readily modify the charge carrier density, giving rise to notable alteration of their optical properties and the resulting forces [18-20]. In addition, marked force contrast has been demonstrated or predicted for configurations based on phase-change materials in their different states [21-26]. However, these techniques usually require non-trivial thermal treatment and the force modulation is not *in-situ*. To date, there have been few attempts at *in-situ* force modulation in response to external stimuli due to experimental difficulties that arise when modulating the optical properties of materials in Casimir measurement configurations. Chen *et al.*, for example, have achieved optical modulation of the Casimir force with up to a few pico-Newtons variation between an Au-coated sphere and a single-crystalline Si membrane through the excitation of charge carriers in the semiconductor using a pulsed Ar laser [27,28]. However, laser-induced Casimir force modulation undergoes undesired artifacts such as heating and exerted optical forces, which can further complicate the experimental consideration.

From the perspective of charge carrier density modulation, electrical biasing/gating is a high-speed modulation technique which is generally easier to operate, less power-consuming and less prone to the above-mentioned artifacts compared to many other techniques. In particular, metal-insulator-semiconductor (MIS) junctions comprising TCOs such as ITO (*i.e.* indium tin oxide) have been widely employed in high-speed electro-optical modulators where the optical responses of the devices can be adapted *on-demand* by tuning the gate voltage, as ITO exhibits gate-controllable optical properties through charge carrier accumulation or depletion at ultrafast speed [29-35]. However, studies on gating-enabled Casimir force modulation are sparse. One recent theoretical work reported the Casimir interaction between a gold platelet and a multilayer stack consisting of a MIS junction made of ITO-Teflon-gold immersed in a liquid environment. It was predicted that the platelet can switch between a "trapped" state and a "released" state by varying the charge carrier density in the ITO layer [36]. Nonetheless, to the best of our knowledge, gate-switchable Casimir forces in an experimentally amenable configuration with pragmatic material and structural parameters and sufficiently measurable force contrast between the "on" and "off" state are still missing.

In this work, we propose two configurations to realize gate-switchable Casimir forces which can be directly deployed in well-established experimental setups. For both configurations the Casimir interaction would be measured between an optically thick Au film-coated sphere, which could be attached to an atomic force microscope (AFM) cantilever for force detection, and a gate-controlled MIS junction consisting of Au-Al$_2$O$_3$-ITO planar films with an applied gate voltage. Two potential configurations are considered (Fig. 1). In configuration I, the ITO film is optically thick and coated by ultrathin layers of Al$_2$O$_3$ and Au. Configuration II is inverted, with a thick film of Au coated with ultrathin layers of Al$_2$O$_3$ and ITO. With a reasonable gate voltage range (0-6 V), the charge carrier density in the ITO accumulation layer can increase by more than an order of magnitude from $10^{19}$ cm$^{-3}$ to $(4-6)\times 10^{20}$ cm$^{-3}$. At short separations (10-50 nm) between the sphere and the MIS stack, the force modulation magnitude is found to reach up to ~15 pN for configuration I and to up to > 400 pN for configuration II, both of which far exceed the measurement sensitivity of the state-of-the-art force measurement techniques using an AFM. Further, we find the thickness of the ultrathin oxide layer between the two electrodes plays a significant role in determining the modulation strength, whose value is enhanced by up to 1.7 times when the thickness is reduced from

3 nm to 2 nm. Our results demonstrate the intriguing prospect of achieving high-speed switchable Casimir forces *in-situ* through electrical gating and provide a rational design for future experimentation.

**Results and discussions**

Figure 1 illustrates the two configurations mentioned above. The radius of the Au-coated sphere is set to be 100 μm, a common value reported in literature [9,18,27,37-42] for Casimir force measurements using an AFM. The Al$_2$O$_3$ layer thickness $t_{ox}$ in the MIS junction is set to be 3 nm, which can be precisely controlled using atomic layer deposition (ALD) technique [43-45]. The top coating layer (Au for configuration I and ITO for configuration II) is set to be thin (5 nm) to warrant a sufficiently large modification to the force while ensuring reasonably good conductivity of the film [31,34,46-48]. In both configurations, the ITO serves as both the active layer for charge carrier density modulation and the electrode for applying gate voltage. Note that the top layer (Au in configuration I and ITO in configuration II) is grounded and therefore a negative bias is applied in configuration I whereas a positive bias is applied in configuration II to the substrate to form the charge accumulation layer at the interface between ITO and the oxide, which is typically 1-3 nm thick [30-34,49,50].

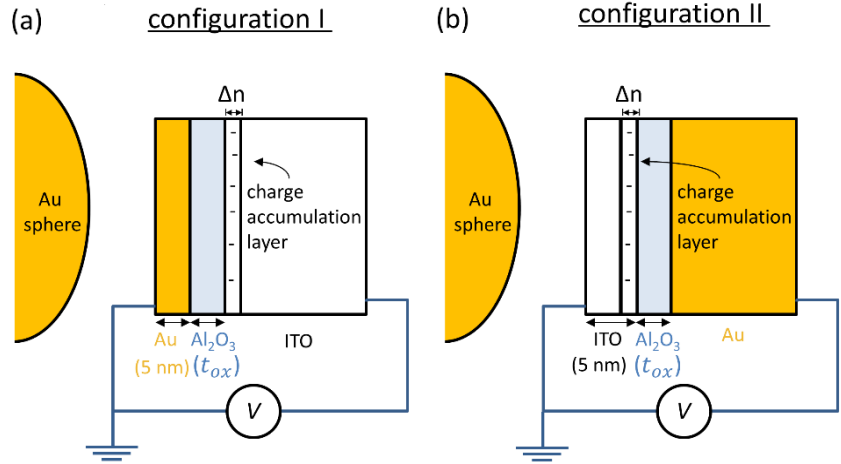

**Figure 1.** Two proposed configurations for actively switchable Casimir forces. A gold sphere of radius $R$=100 μm is brought close to a metal-insulator-semiconductor (MIS) junction consisting of an Au layer and an ITO layer sandwiching an Al$_2$O$_3$ ultrathin film. When the junction is gated, the charge carrier density at the interface between the oxide and ITO is significantly increased, forming an ultrathin accumulation layer with modified optical properties compared to the otherwise as-deposited ITO film due to the charge accumulation. The Casimir force between the Au sphere and the MIS junction is thus modified. The orientation of the MIS junction is different for the two configurations, as (a) the Au film faces the Au sphere and (b) the ITO layer faces the Au sphere.

To quantify the charge accumulation effect at the ITO-oxide interface, we utilize a simple capacitance model across an MIS junction which assumes a uniform carrier density in the ultrathin accumulation layer, as widely adopted in literature [30,35,48-50]. The average thickness $t_{acc}$ of the accumulation layer due to carrier injection in a standard MIS junction is given by [51]: $t_{acc} = \frac{\pi}{\sqrt{2}}\sqrt{\frac{k_B T \varepsilon_0 \varepsilon_S}{N_0 q^2}}$, where $k_B$ is the Boltzmann constant, $T = 300$ K is the room temperature, $\varepsilon_0$ is the free-space permittivity, $\varepsilon_S = 9.3$ is the relative static permittivity of ITO [30,32,34], $q$ is the electron charge, and $N_0$ is the initial carrier density in the ITO layer. In practice, the carrier density in ITO as-deposited is dependent upon the deposition processes and annealing conditions [52,53], thus can vary by as large a range as $10^{19}$-$10^{21}$ cm$^{-3}$. We set the ITO initial carrier density as $1\times10^{19}$ cm$^{-3}$ for our computation, the same as reported in literature [34,35,49], which yields $t_{acc} = 2.56$ nm. When a gate voltage $V_g$ is applied across the junction, the carrier density in the accumulation layer can be written as:

$$N_{acc} = N_0 + \frac{\varepsilon_0 \varepsilon_{ox} V_g}{q t_{ox} t_{acc}} \quad (1)$$

where $\varepsilon_{ox} = 9$ denotes the relative static permittivity of Al$_2$O$_3$ [32,48,54]. Here we restrict $V_g < 6$ V to avoid electrical breakdown of the oxide [44,45,55], which increases the carrier density in the accumulation layer to about $4\times10^{20}$ cm$^{-3}$, more than an order of magnitude larger than the initial value. Such a profound modulation of carrier density *via* gating in ITO-based MIS junctions has also been reported by a number of experimental works [29,34,35,49,56].

The Casimir force between a sphere with radius $R$ and a planar structure at a separation $d$ is given by the Lifshitz formula using proximity force approximation (PFA) provided $R \gg d$ [3,57]:

$$F(d) = k_B T R \sum_{m=0}^{\infty}{'} \int_0^{\infty} k [\ln(1 - r_1^{TE} r_2^{TE} \exp(-2k_\perp d)) + \ln(1 - r_1^{TM} r_2^{TM} \exp(-2k_\perp d))] dk, \quad (2)$$

where $k$ is the lateral wavenumber, $k_\perp = \sqrt{k^2 + \frac{\xi_m^2}{c^2}}$ is the vertical wavenumber in the intervening medium (air), $\xi_m = \frac{2\pi k_B T}{\hbar} m$ denotes the Matsubara frequencies, the prime sign on the summation means the zero-frequency term is multiplied by half, and $r_i^\sigma$ ($i = 1,2$ and $\sigma = $ TE, TM) represent the reflection coefficients at the interface between air and medium $i$ (note: the Au sphere is medium 1 and the MIS stack is medium 2 for our configuration) for imaginary frequency $\xi_m$ and lateral wavenumber $k$ under TE and TM polarizations. The reflection coefficients off the surface of the stack are computed using the transfer matrix method (TMM) [36]. Because the reflection depends naturally on the material's broadband dispersion (dielectric function) and the object's geometry and size, so does the resulting force.

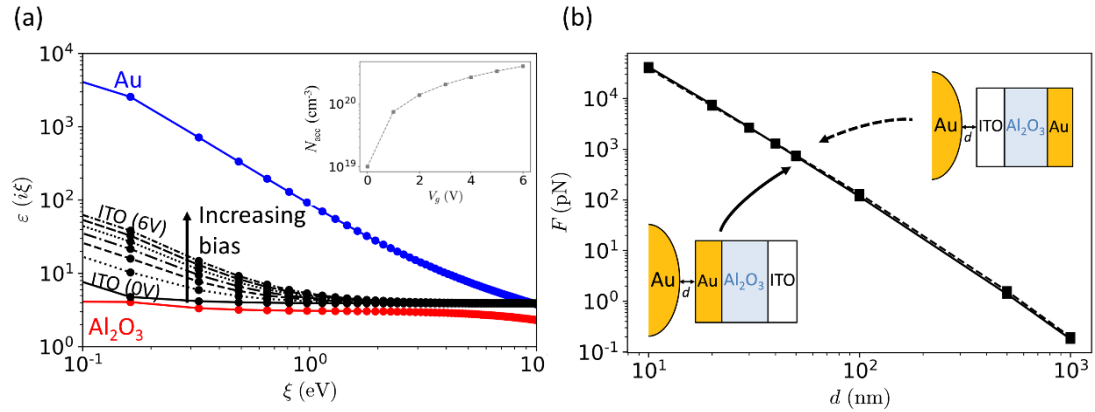

**Figure 2.** (a) Dielectric functions for different materials with respect to the Matsubara frequencies at room temperature. The dielectric function of the ITO accumulation layer monotonically increases with applied gate voltage (oxide thickness is 3 nm in the MIS junction). Inset shows the carrier density increase in the accumulation

layer in ITO with applied gate voltage. (b) Casimir force between the sphere and the MIS stack under zero gate voltage as a function of separation. The solid and dotted line represent the force for configuration I and II, respectively.

We apply dielectric function data/models for the materials using the most often utilized data for Casimir force calculations. The optical data for Au is obtained from Palik's handbook, extended to lower energies using a Drude model with parameters $\omega_p = 9$ eV and $\gamma_p = 0.035$ eV [8,9,18,41,58,59]. The dielectric function for $Al_2O_3$ is modeled using a dual-oscillator Lorentz model [7,60]. For ITO, we apply a Drude model with the following parameters [32,34-36,49,50]: $\varepsilon_\infty = 3.9$, $\gamma_p = 1.8 \times 10^{14}$ s$^{-1}$, and $\omega_p = \sqrt{Nq^2/\varepsilon_0 m^*}$, where $N$ is the charge carrier density and $m^* = 0.35 m_e$ is the charge carrier effective mass, with $m_e$ being the electron mass. Figure 2a shows the dielectric functions of the abovementioned materials. As expected, the permittivity values with respect to Matsubara frequencies for the accumulation layer in ITO lie between those for Au and $Al_2O_3$ and monotonically increase with applied gate voltage as a result of augmented carrier density, which renders the interface more "metallic". The calculated Casimir forces for both configurations under zero gate voltage are shown in Fig. 2(b). We note that they exhibit commensurate force magnitudes in this separation range.

When a gate voltage is applied across the junction, the force magnitude is modified due to the change of charge carrier density in the accumulation layer (Fig. 3), which ultimately alters the overall reflection at the top surface of the stack. The force modulation $\Delta F$ (compared with zero gate voltage) is over an order of magnitude larger for configuration II compared to configuration I. Further, we find that the force modulation reaches > 400 pN when the separation is reduced to 10 nm with an applied bias of 6 V. Contrastingly, the force modulation is much less than 1 pN with separations greater than 50 nm. Note that the positive values for $\Delta F$ means the force becomes more attractive when gate voltage is turned on, in agreement with the intuition that the stack becomes more metallic as a result of charge carrier injection.

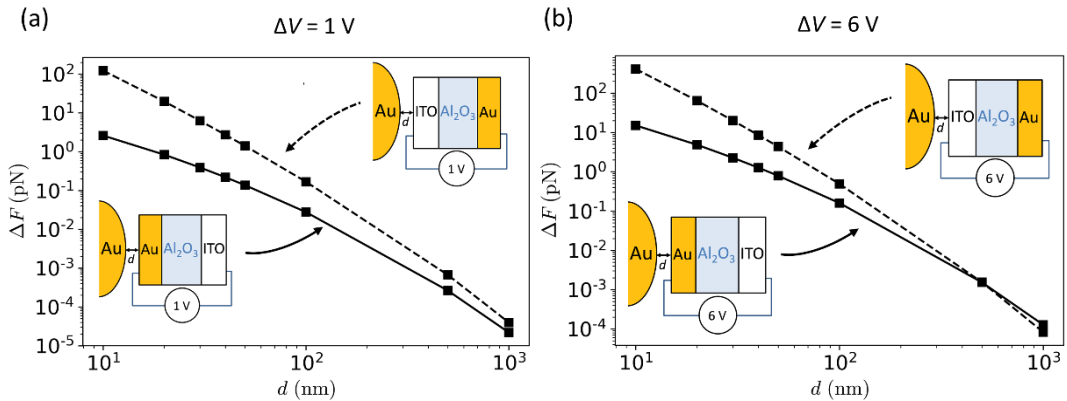

**Figure 3.** Force modulation as a function of separation under the gate voltage of (a) 1 V and (b) 6 V. The force change in configuration II is on average more than one order of magnitude larger than in configuration I. The solid and dotted line represent the force modulation for configuration I and II, respectively.

From the experimental point of view, state-of-the-art AFM techniques with a sphere-planar configuration feature an optimal force measurement sensitivity of 1-2 pN [8,18,27,41,61,62]. This indicates that to obtain a measurable force modulation, a separation of less than 30 nm is needed while the gate voltage is switched between 0 and 6 V. Despite the difficulty due to the jump-to-contact (JTC) effect and surface roughness to consider, this short separation range is achievable in experiments [7,21,40,63,64]. Further, since the force modulation is generated by switching the gate voltage on and off, it can, in principle, be directly measured with a better sensitivity using a lock-in amplifier by referencing the voltage on-off control signal at a particular modulation frequency (~100-1000 Hz) for phase-locking. In fact, Chen *et al.* employed a similar technique to measure the laser-induced force modulation, reducing the measurement noise to the level of 0.1-0.5 pN [27,28].

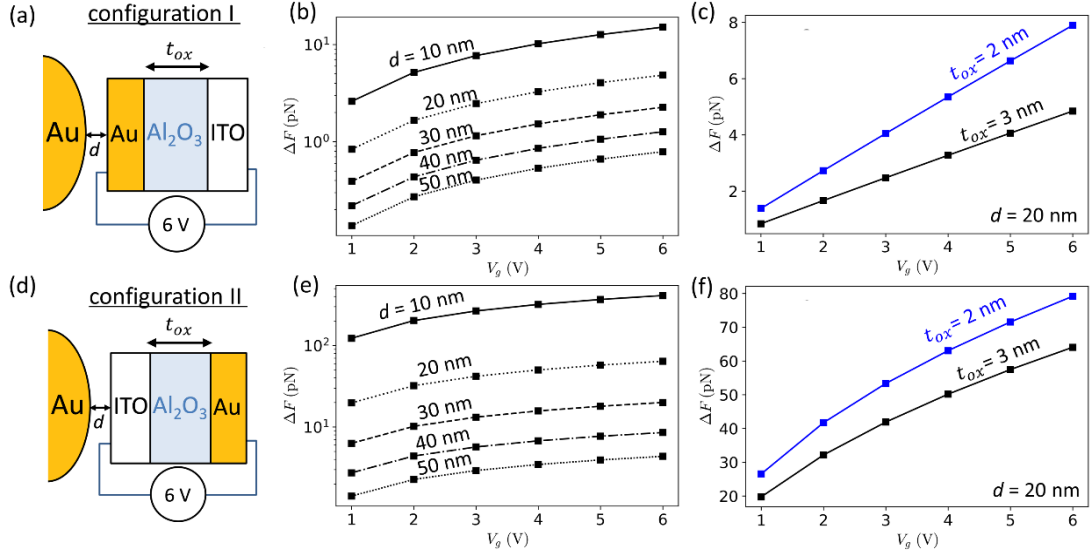

**Figure 4.** Modulation of the Casimir force with applied gate voltage and oxide layer thickness. (a) Schematic of configuration I showing (b) the force change at different separations with an oxide thickness of 3 nm in the MIS junction. (c) Force change at a fixed separation of 20 nm, with two different oxide thicknesses (2 nm and 3 nm, respectively). (d,e,f) Same as (a-c) but for configuration II.

At a fixed separation, the force modulation varies as a function of both the applied voltage bias and insulating layer ($Al_2O_3$) thickness (Fig. 4). We find similar behavior for configurations I and II (Fig. 4a and Fig. 4d, respectively), with more pronounced variations for configuration II. For both configurations, $\Delta F$ monotonically increases with increasing gate voltage due to the enhanced reflection of the structure with increased carrier density. Figure 4c shows how the force modulation is controlled by the gate voltage at a separation of 20 nm with two different oxide thicknesses. The reduction of the thickness from 3 nm to 2 nm enhances the modulation magnitude by more than 60%, resulting in $\Delta F \sim 8$ pN for a 6 V gate voltage. The strong dependence of the force change on the oxide thickness is attributed to the change of carrier accumulation at the ITO-oxide interface. With a 2-nm thick oxide layer, the carrier density reaches $5.93 \times 10^{20}$ cm$^{-3}$ with a 6 V gate voltage, about 1.5 times that for a 3-nm oxide layer. One caveat of utilizing a thinner oxide layer is that the maximum gate voltage to be applied is further constrained by the breakdown field strength of the oxide. Fortunately, precise control of the oxide layer at the level of sub-nanometer scale is made possible by advanced deposition techniques such as ALD [55].

Compared with configuration I, the modulation for configuration II (Fig. 4(d)-(f)) is on average one order of magnitude stronger in the separation range we considered, which allows for a measurable force modulation even with just 1 V gate voltage switched on and off. This behavior can be ascribed to the closer distance between the charge accumulation layer and the Au sphere, leading to a much greater reflection change at the top surface of the stack. Because the force modulation magnitude is significantly greater for this configuration, we anticipate that configuration II will be much easier to embody in experiment. Likewise, reduction of the oxide thickness to 2 nm increases $\Delta F$ by 1.2-1.3 times, as shown in Figure 4f. One interesting visual distinction between the two configurations is how $\Delta F$ scales with the gate voltage $V_g$ (and the resulted variation of the carrier density $N_{acc}$). In configuration I, $\Delta F$ increases almost linearly with $V_g$. Contrastingly, the variation of $\Delta F$ with $V_g$ is more nonlinear in configuration II. Nonetheless, the visually perceived linearity for configuration I is merely the result of very small force modulation values. This behavior is another manifestation of the highly complex nature of the relation between material's local optical properties and the force.

**Conclusion**

In summary, we theoretically investigated two configurations for potential implementation of gate-switchable Casimir forces, both of which are composed of a gate-controlled MIS junction of Au-$Al_2O_3$-ITO planar films, with different orientations towards a gold-coated sphere attached to an AFM cantilever. The charge carrier density in the ITO accumulation layer formed at the interface between ITO and the oxide layer can be tuned substantially from $10^{19}$ cm$^{-3}$ to (4-6)×$10^{20}$ cm$^{-3}$ *via* gating. As a result, a force modulation magnitude reaches up to > 400 pN with a gate voltage of 6 V, far exceeding the measurement sensitivity with the state-of-the-art AFM force-measurement techniques. Furthermore, a reduction of the oxide layer thickness from 3 nm to 2 nm can increment the force modulation magnitude by up to 70%, which indicates the precise control of the oxide layer thickness *via* advanced deposition techniques such as ALD is paramount for force modulation. Our results show the great promise of utilizing TCO materials to realize switchable Casimir forces with a pronounced force contrast, which may create new opportunities for *in-situ* control and modulation of movable parts in nanomechanical devices and systems.

**Acknowledgements**

The authors acknowledge financial support from the Defense Advanced Research Program Agency (DARPA) QUEST program No. HR00112090084.